\documentclass[12pt,preprint]{aastex}

\newcommand{\swift}{\mbox{\em Swift}}

\newcommand{\dex}[1]{\hbox{$\times\hbox{10}^{#1}$}}

\newcommand{\kmsec}{\,\mbox{km\,s$^{-1}$}}

\newcommand{\nh}{\,\mbox{$N_{\mathrm H}$}}
\newcommand{\betauvot}{\,\mbox{$\beta_{\mbox{\tiny UVOT}}$}}
\newcommand{\betaxrt}{\,\mbox{$\beta_{\mbox{\tiny XRT}}$}}
\newcommand{\betabat}{\,\mbox{$\beta_{\mbox{\tiny BAT}}$}}
\newcommand{\alphauvot}{\,\mbox{$\alpha_{\mbox{\tiny UVOT}}$}}
\newcommand{\alphaxrt}{\,\mbox{$\alpha_{\mbox{\tiny XRT}}$}}

{\newcommand{\etal}{\mbox{et\ al.\ }}

\slugcomment{Accepted for publication in Astrophysical Journal Letters}

\shorttitle{{\em Swift}-UVOT detection of GRB 050318}
\shortauthors{M. Still \etal}

\begin{document}

\title{{\em Swift}-UVOT detection of GRB 050318}

\author{M. Still\altaffilmark{1,2}, 
P.W.A. Roming\altaffilmark{3}, 
K.O. Mason\altaffilmark{4} 
A. Blustin\altaffilmark{4}, 
P. Boyd\altaffilmark{1,5}, 
A. Breeveld\altaffilmark{4}, 
P. Brown\altaffilmark{3}, 
M. De Pasquale\altaffilmark{4},
C. Gronwall\altaffilmark{3},  
S.T. Holland\altaffilmark{1,2}, 
S. Hunsberger\altaffilmark{3}, 
M. Ivanushkina\altaffilmark{2,6}, 
C. James\altaffilmark{4},
W. Landsman\altaffilmark{1}, 
K. McGowan\altaffilmark{4}, 
A. Morgan\altaffilmark{3},
T. Poole\altaffilmark{4}, 
S. Rosen\altaffilmark{4}, 
P. Schady\altaffilmark{3,4}, 
B. Zhang\altaffilmark{7}, 
H. Krimm\altaffilmark{1,2}, 
T. Sakamoto\altaffilmark{1},
P. Giommi\altaffilmark{8},
M.R. Goad\altaffilmark{9},
V. Mangano\altaffilmark{10},
K. Page\altaffilmark{9}, 
M. Perri\altaffilmark{8},
D.N. Burrows\altaffilmark{3},
N. Gehrels\altaffilmark{1},
J. Nousek\altaffilmark{3}} 
\affil{}

\altaffiltext{1}{NASA Goddard Space Flight Center, Greenbelt, MD 20771}
\altaffiltext{2}{Universities Space Research Association}
\altaffiltext{3}{Department of Astronomy and Astrophysics, Pennsylvania State University, 525 Davey Laboratory, University Park, PA 16802}
\altaffiltext{4}{Mullard Space Science Laboratory, University College London, Holmbury St. Mary, Dorking, RH5 6NT Surrey, UK}
\altaffiltext{5}{Joint Center for Astrophysics, University of Maryland, Baltimore County, 1000 Hilltop Circle, Baltimore, MD 21250}
\altaffiltext{6}{Department of Physics and Astronomy, Brigham Young University, N208 ESC, Provo, UT 84602}
\altaffiltext{7}{Department of Physics, University of Nevada, 4505 Maryland Parkway, Las Vegas NV 89154}
\altaffiltext{8}{ASI Science Data Center, ESRIN, 00044 Frascati, Italy}
\altaffiltext{9}{Department of Physics \& Astronomy, University of Leicester, Leicester LE1 7RH, UK}
\altaffiltext{10}{Osservatorio Astronomico di Roma, Via Frascati 33, 00040 Monteporzio Catone (Roma), Italy}

\begin{abstract} 

We present observations of GRB 050318 by the Ultra-Violet and Optical
Telescope (UVOT) on-board the \swift\ observatory.  The data are the
first detections of a Gamma Ray Burst (GRB) afterglow decay by the
UVOT instrument, launched specifically to open a new window on these
transient sources.  We showcase UVOTs ability to provide multi-color
photometry and the advantages of combining UVOT data with simultaneous
and contemporaneous observations from the high-energy detectors on the
\swift\ spacecraft.  Multiple filters covering
$\lambda\lambda$1,800--6,000\AA\ reveal a red source with spectral
slope steeper than the simultaneous X-ray continuum. Spectral fits
indicate that the UVOT colors are consistent with dust extinction by
systems at $z = 1.2037$ and $z = 1.4436$, redshifts where absorption
systems have been pre-identified. However, the data can be most-easily
reproduced with models containing a foreground system of neutral gas
redshifted by $z = 2.8 \pm 0.3$. For both of the above scenarios,
spectral and decay slopes are, for the most part, consistent with
fireball expansion into a uniform medium, provided a cooling break
occurs between the energy ranges of the UVOT and \swift's X-ray
instrumentation.

\end{abstract}


\section{Introduction}
\label{sec:introduction}

The multi-instrument \swift\ observatory (Gehrels \etal 2004) was
launched on Nov 20, 2004.  It carries three science instruments, the
wide-angle, hard X-ray, Burst Alert Telescope (BAT; Barthelmy \etal
2005) which locates GRBs to within $3^\prime$ on the sky, the
narrow-field X-Ray Telescope (XRT; Burrows \etal 2005) and the UVOT.
Specifications of the UVOT are described in Roming \etal (2005).  The
UVOT instrument has a vital role imaging the field containing the
burst, minutes after a trigger, and reporting rapidly the afterglow
location to $<1^{\prime\prime}$ accuracy via the GRB Coordinate
Network (GCN).  UVOTs subsequent role is to provide a
relatively-uniform sample of the afterglow decay.  It is this
subsequent role that we report on here, describing the first afterglow
detected by UVOT in multiple colors, and monitoring the decay until 40
ksec after the burst.

\section{Observations} 
\label{sec:observations}

The \swift-BAT made a 17$\sigma$ detection of GRB 050318 at 15:44:37
UT (Krimm \etal 2005a).  Burst parameters revised from Krimm \etal
(2005c) include a $T_{\mathrm 90}$ burst duration of $32 \pm 2$ s,
with a total fluence of $2.1 \dex{-6}$ erg cm$^{-2}$ in the 15--350
keV band. Within this energy band we find evidence of spectral
evolution across three peaks in the prompt emission light curve.  Peak
1, between $T$--1s and $T$+5s, (where $T$ is the trigger time), is
well fit by a simple power law with spectral index \betabat\ = $-1.1
\pm 0.2$ ($\chi^2 = 57$ for 57 d.o.f.).  All uncertainties in this
paper are reported to a 90\% confidence level while spectral and
temporal decay indices are provided with respect to flux density,
e.g., $F_\nu \sim t^{\alpha}\nu^{\beta}$.  The burst was quiet for the
next 17 seconds, followed by two overlapping, but resolved peaks.
Peak 2 ($T$+22--27s) fits to a cut-off power law with \betabat\ =
$-0.2 \pm 0.5$ and $E_{\mathrm{p}} = 68^{+23}_{-10}$ keV ($\chi^2 =
66$ for 56 d.o.f.).  The spectrum softens considerably during the
third peak ($T$+27--32s), and is fit with \betabat\ = $0.2 \pm 0.4$,
where $E_{\mathrm{p}} = 46 \pm 7$ keV.  BAT event data were not
recorded for the final 2s of the burst.  However, examination of the
BAT rate data in four energy bands suggests continued spectral
softening during this period.

The burst was located to within $3^\prime$ (90\% containment) of RA =
49.651, Dec = $-46.392$ (J2000).  This corresponds to a Galactic
latitude of $-55^{\circ}$ with a local reddening of $E$(B-V) = 0.018
mag (Schlegel, Finkbeiner \& Davis 1998) and a H-equivalent Galactic
column density of $\nh = 2.8 \dex{20}$ cm$^{-2}$ (Dickey \& Lockman
1990).  After a 54 min delay for Earth occultation, \swift\
slewed so that the narrow-field instruments could monitor the target.

\begin{figure}                          
\begin{picture}(100,0)(10,20)
\put(0,0){\includegraphics{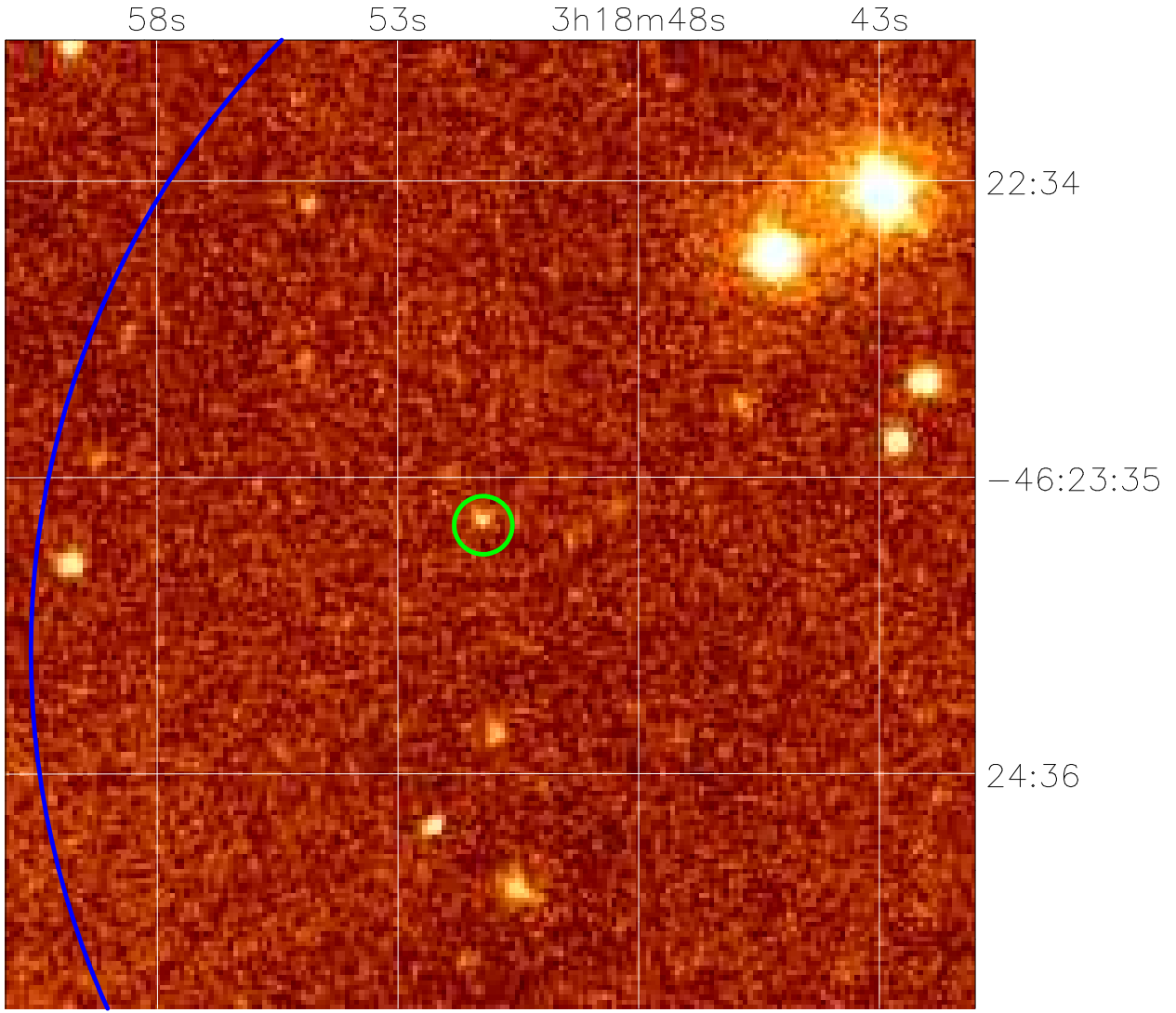}}   
\noindent    
\end{picture}    
\vspace{105mm} 
\figcaption{Stacked UVOT-V filter image of the field
  with the transient source at RA = 03$^{\mathrm h}$ 18$^{\mathrm m}$
  51$^{\mathrm s}$.15, Dec = -46$^{\circ}$ 23$^{\prime}$
  43$^{\prime\prime}$.7 (J2000) and $3^\prime$ BAT error circle and
  $6^{\prime\prime}$ XRT error circles overlaid.  Total exposure time
  for the stacked image is 3,732s.\label{image}}
\end{figure}

Within the first 100s settled observation of the UVOT sequence, a V =
17.8 source was found $2^{\prime}.6$ from the BAT position
(Fig~\ref{image}), with no counterpart in archival plates (McGowan
\etal 2005) and consistent with the ground-based report of Mulchaey \&
Berger (2005).  Subsequent exposures revealed a fading source,
$1^{\prime\prime}.1$ from a transient X-ray counterpart (Markwardt
\etal 2005; Nousek \etal 2005; Beardmore \etal 2005).  A complete
analysis and description of the XRT data reduction is reported in a
separate paper (Perri \etal 2005).  The UVOT position, as reported by
De Pasquale \etal (2005), is RA = 03$^{\mathrm h}$ 18$^{\mathrm m}$
51$^{\mathrm s}$.15, Dec = -46$^{\circ}$ 23$^{\prime}$
43$^{\prime\prime}$.7 (J2000), with $0^{\prime\prime}.3$
uncertainties.

The UVOT completed 36 exposures before GRB 050319 triggered the BAT
and became the new automated target (Krimm \etal 2005b).  All
detections of GRB 050318 $\geq 2\sigma$ above the background are
tabulated in Table~\ref{tab:mags}.  AB filter magnitudes and
background limits are based upon in-orbit zero-point calibrations, and
differ from those used by McGowan \etal (2005) and De Pasquale \etal
(2005), which were based on pre-flight calibrations and Vega
magnitudes.  The afterglow is not detected in UVW1 (centered at
approximately 2,500\AA), UVM2 (2,200\AA) or UVW2 (1,800\AA) light; the
first settled exposures yield $3\sigma$ upper limits of 19.3, 19.5 and
21.2 mag, respectively.  Detections are made through the U (3,500\AA),
B (4.400\AA) and V (5,300\AA) filters between T+3,200--5,400s. On the
next rotation of the filter wheel at $\sim$ T+21,000s the source has
decayed below the $3\sigma$ background threshold in both the U and B
bands.  The magnitudes and detection significances at this epoch are U
= $21.9 \pm 0.5$, detected at $2.3\sigma$ above background and B =
$21.6 \pm 0.8$, $1.3\sigma$ above background.  The V source persists
for three wheel rotations before fading below the background threshold
between T+23,000--34,000s. The U band contains two further marginal
source detections between 2--3$\sigma$ above background at T+28,609
and T+40,193.

\subsection{Source decay}
\label{sec:decay}

\begin{figure}
\begin{picture}(100,0)(10,20) 
\put(0,0){\includegraphics{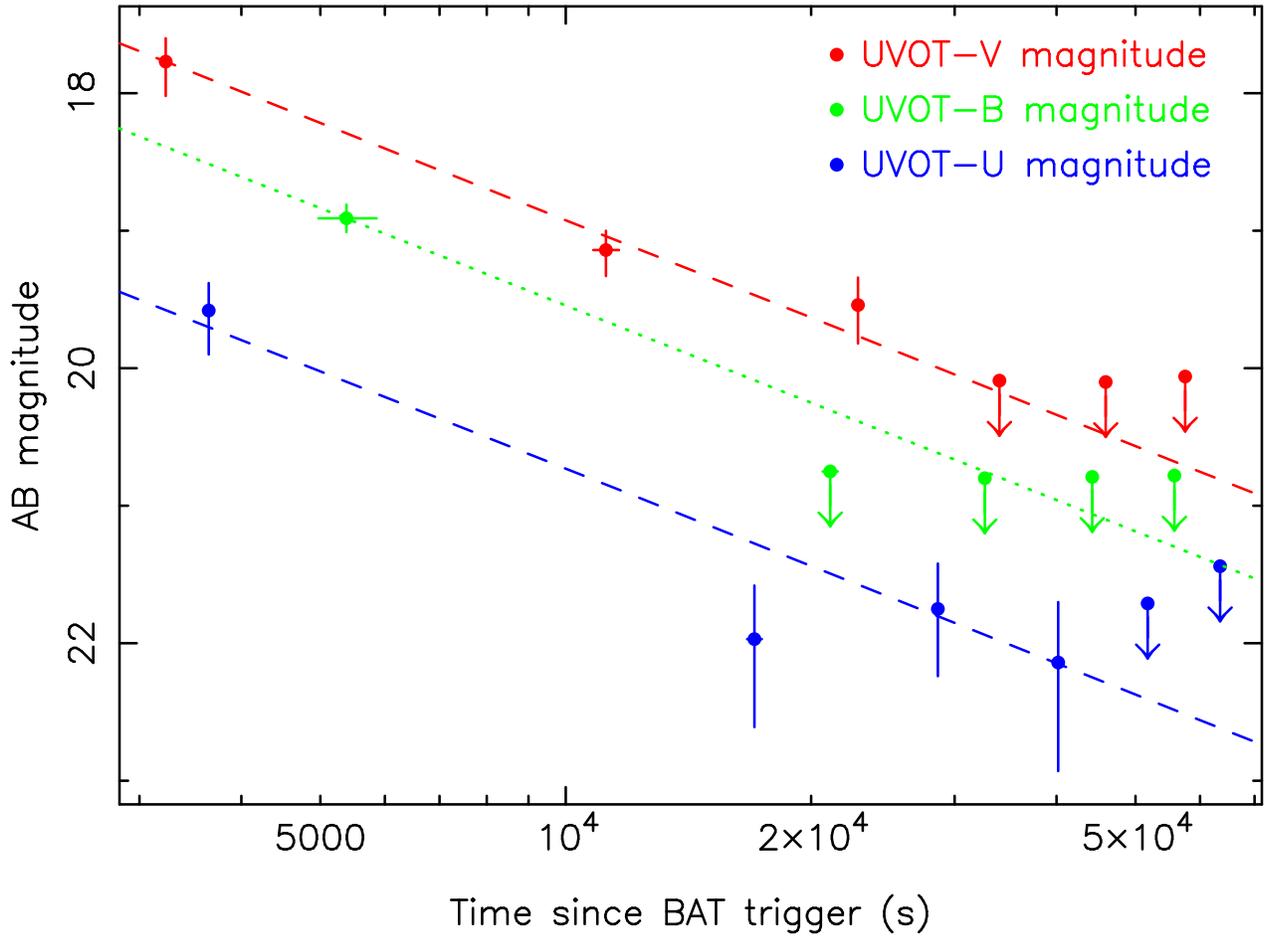}}
\noindent
\end{picture}
\vspace{100mm}
\figcaption{U,   B and   V   light
curves of GRB 050318.  The dashed lines are the   best powerlaw fits to  the
U and V  time-series, $\alpha_{\mbox{\tiny U+V}}$, excluding upper-limits.
The Dotted  curve   is an identical powerlaw model,
renormalized  to   the  first-epoch B magnitude.\label{lc}} 
\end{figure}

All U, B and V points, bar the first 2s B exposure, are plotted in
Fig.~\ref{lc}.  Detections $\geq 2\sigma$ are provided with 90\%
confidence error bars, while all other points are given as upper
limits at the $3\sigma$ level.  Using the first three V filter
exposures, the powerlaw decay index for the V light curve is
$\alpha_{\mathrm V} = -0.87 \pm 0.24$ ($\chi^2 = 1.1$ for 1 dof).
Assuming a powerlaw decay, the four U band detections, $\geq 2\sigma$
above background, yield a consistent slope $\alpha_{\mathrm U} = -1.00
\pm 0.25$ (90\% confidence, $\chi^2 = 2$ for 2 dof).  A weighted-mean
of the U and V decay slopes provides $\alpha_{\mathrm U+V} = -0.94 \pm
0.17$ and the best fits to U and V data using this slope are plotted
in Fig.~\ref{lc}. A curve of the same slope is extrapolated to pass
through the B detection at T+5,382s. The XRT light curve at this
epoch, presented by Perri \etal (2005), has a powerlaw decay index of
$\alpha_{\mbox{\tiny XRT}} = -1.2 \pm 0.1$. The probability that
$\alpha_{\mbox{\tiny XRT}}$ and $\alpha_{\mbox{\tiny U+V}}$ are
identical is 8\%.

Source detection in the T+21,105s B exposure is significant only to
$1.3\sigma$ and, using the B detection at T+5,382s as an anchor point,
is inconsistent with the powerlaw decay index of $\alpha_{\mathrm
  U+V}$ with 99.9\% confidence. Either there is under-sampled
variability in the source which provides us with a biased measure of
the decay indices, or we are observing spectral evolution. Perhaps it
is no coincidence that the next exposure after the second B band
observation is the one point on the U curve that is an outlier
relative to the best-fit powerlaw decay index.  It is inconsistent
with $\alpha_{\mathrm U+V}$ with 96\% confidence.  So short-timescale
variability is perhaps the most plausible interpretation.


\subsection{Spectral properties}
\label{sec:spectrum}

The two absorption systems at $z_1 = 1.2037$ and $z_2 = 1.4436$
reported by Berger \etal (2005) should produce Lyman systems
redshifted into the UVM2 band. Assuming that consistent non-detections
in filters blueward of a particular wavelength reveal either dust or
the Lyman limit of the host and its redshift, in this section we
formally measure the spectral slope across the UVOT bands and search
for a Lyman edge.  All spectral models below include Galactic
extinction appropriate for the source direction
(Sec.~\ref{sec:introduction}), using the analytic formalism of Pei
(1992) with $R_V = 3.08$. For simplicity, we assume that there is no
Ly\,$\alpha$ forest in front of the host.

Using $\alpha_{\mbox{\tiny U+V}}$, Source rates in each filter were
interpolated or extrapolated to a common epoch of T+4,061s. Using the
$\chi^2$ fitting method outlined in Arnaud (1996) and references
therein, a simple powerlaw model to all six points yields a fit with
spectral index \betauvot\ = $-4.9 \pm 0.5$ and $\chi^2$ = 24 for 4
dof. The fit is poor statistically and the index is steep compared to
the 0.2--5 keV slope obtained from simultaneous XRT data of
\betaxrt\ = $-1.1^{+0.2}_{-0.4}$ (Sec.~\ref{sec:sed}).  By adding an
absorption edge to the powerlaw model, the quality of the fit improves
to a statistically acceptable solution; \betauvot\ = $-2.4 \pm 1.5$
and, assuming the edge is due to the Lyman series, $z = 2.8 \pm 0.2$
($\chi^2$ = 1.0 for 3 dof). The best-fit redshift is inconsistent with
$z_2$ = 1.4436 (from Berger \etal 2005). For comparison, if $z$ is
forced to be 1.4436, the best fit yields \betauvot\ = $-4.5 \pm 1.5$
with $\chi^2$ = 15 for 4 dof. For completeness, fixing $z$ at a value
of 1.2037 yields \betauvot\ = $-4.5 \pm 0.5$ ($\chi^2$ = 20 for 4
dof).

We add host extinction to the model above, assuming an SMC grain
content with $R_V = 2.93$ (Pei 1992), and coupling the redshift of the
dust to the neutral gas.  The best solution yields \betauvot\ = $-1.0
\pm 5.0$, $z = 2.9 \pm 0.2$ and $E$(B-V) $< 0.26$ ($\chi^2$ = 1.0 for
2 dof), i.e. dust is not a necessary component for this model in order
to provide a good description of the UVOT data. However, when the dust
and gas is re-situated at $z_2$, the fit does converge to an
acceptable solution, provided we also include dust and gas at $z_1$,
with $\chi^2$ = 3.2 for 2 dof, \betauvot\ = $+1.0 \pm 2.0$,
$E$(B-V)$_1 = 0.4 \pm 0.2$ and $E$(B-V)$_2 < 0.27$, where $E$(B-V)$_1$
and $E$(B-V)$_2$ are the color excesses at $z_1$ and $z_2$
respectively.  Consequently, gas and dust, at the redshifts of the two
absorption systems reported by Berger \etal (2005) in front of a
powerlaw continuum provide an adequate fit to the UVOT data. However,
the model containing a single gas and dust complex at the larger
redshift of $z = 2.8 \pm 0.2$ provides the better fit.

Next we investigate whether the solutions above are biased by assuming
an inappropriate temporal decay slope during the interpolation of
UVOT data to a common epoch. For comparison, we repeat the previous
exercise using the XRT decay index \alphaxrt\ = $-1.2$. Best-fit
parameters and fit quality vary only a little compared to the previous
analysis, and this results from the choice of a common epoch which
minimizes the systematic uncertainty in the interpolation. The
best-fit solution without a dust component in the host (which does not
formally improve the fit) is \betauvot\ = $-1.0 \pm 5.0$ and $z = 2.9
\pm 0.5$ ($\chi^2$ = 1.0 for 3 dof). The best fit with two dust and
gas systems at $z_1$ and $z_2$ yields $\chi^2 = 2.6$ for 2 dof,
\betauvot\ = $+1.0 \pm 4.1$, $E$(B-V)$_1 = 0.5 \pm 0.4$ and
$E$(B-V)$_2 < 0.35$. Both solutions are identical to the previous
analysis within uncertainties.

Intrinsic continuum slopes in the above models are poorly constrained
due to a combination of low count rates and the relatively small
spectral range of the filters. In the next sections, by combining the
UVOT data with a simultaneous XRT spectrum, we can place further
constraints on the UVOT continuum, refine the redshift test and dust
measurements, and compare a simple fireball model to the data.

\subsection{Spectral energy density}
\label{sec:sed}

Good XRT events have been extracted from within the time interval
T+3,180--5,822s, which is the epoch between the start of the first V
exposure and the end of the subsequent B detection, and binned by
pulse height. The spectral fits below contain a core model of a
floating powerlaw, combined with fixed quantities for Galactic
reddening and extinction in the local rest frame
(Sec.~\ref{sec:introduction}). Galactic abundances are from Anders \&
Grevesse (1989).  We use the April 5, 2005 empirical version of the
XRT response calibration, which requires an additional absorption
feature added to spectral models, corresponding to the neutral O K
feature at 0.54 keV, due to the optical filter.

The best fit to the XRT data alone yields a powerlaw slope of
\betaxrt\ = $-1.2 \pm 0.3$ and an integrated 0.2--5 keV flux of $(1.7
\pm 0.3) \dex{-11}$ erg s$^{-1}$ cm$^{-2}$.  No extra spectral
components are required with $\chi^2 = 10$ for 21 dof.  On comparing
the UVOT spectrum to the XRT spectral model, not only do we find an
observed optical/UV spectral index much steeper than the X-ray
continuum, but also the UV fluxes are $>1$ order of magnitude fainter
than those predicted by the XRT model, therefore the different slopes
cannot be caused by a spectral break alone. A combined fit of the UVOT
and XRT data to the core model yields a poor fit with a spectral index
$\beta = -0.45 \pm 0.03$ and $\chi^2$ = 506 for 28 dof.

\begin{figure}
\begin{picture}(100,0)(10,20) 
\put(0,0){\includegraphics{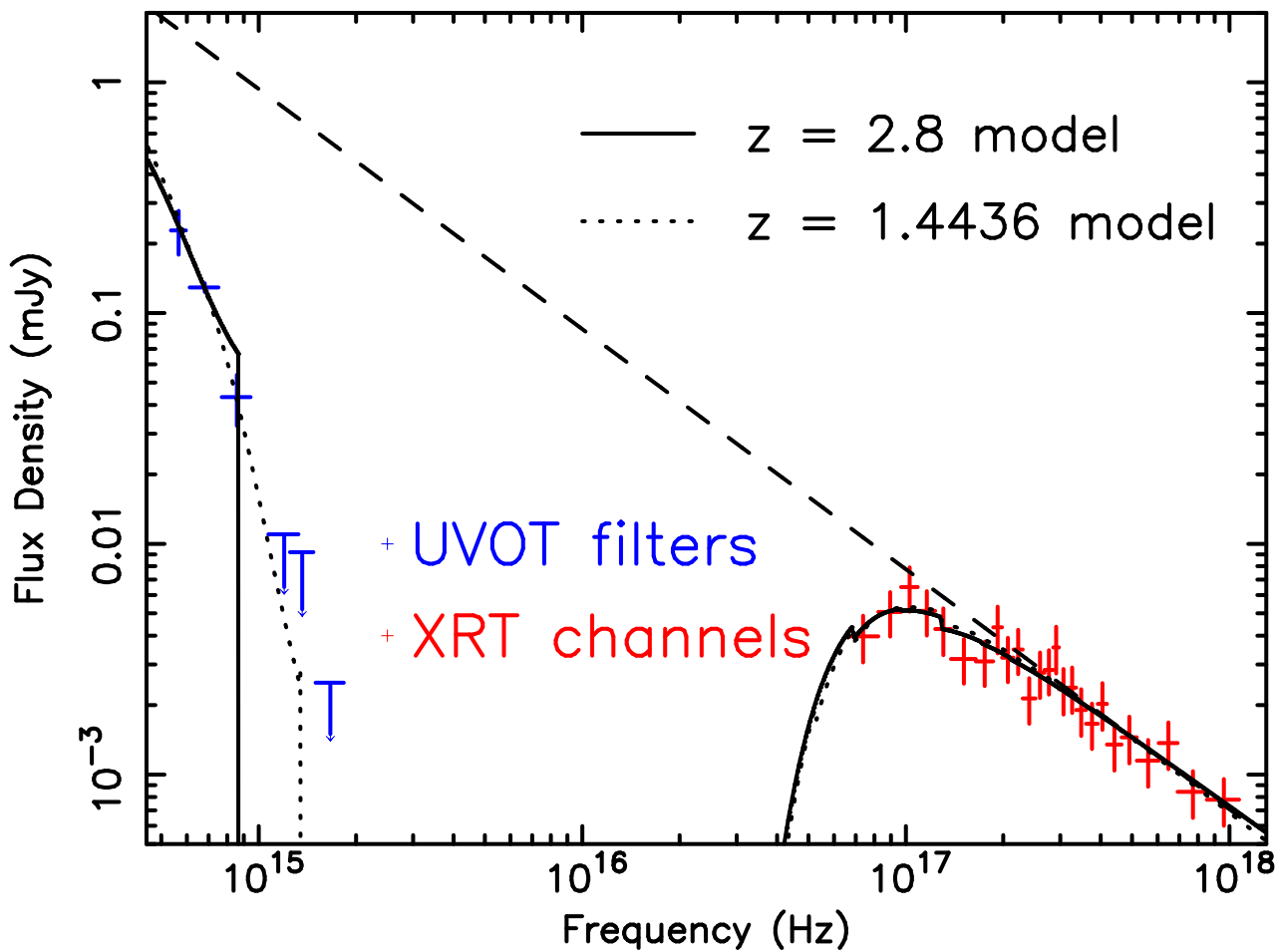}}
\noindent
\end{picture}
\vspace{105mm} 
\figcaption{The combined UVOT and XRT spectrum of GRB
  050318 at epoch T+4,061s, compared to two best-fit models, both
  containing a powerlaw model, reddened and absorbed by Galactic
  material. The Solid line represents this model with an additional
  system of neutral gas and SMC-like dust at $z = 2.8$. The dotted
  line represents the model with two systems of neutral gas and
  SMC-like dust at $z_1 = 1.2037$ and $z_2 = 1.4436$. The dashed line
  is the best-fit intrinsic powerlaw spectrum.
\label{sed}} 
\end{figure}

\begin{figure}
\begin{picture}(100,0)(10,20) 
\put(0,0){\includegraphics{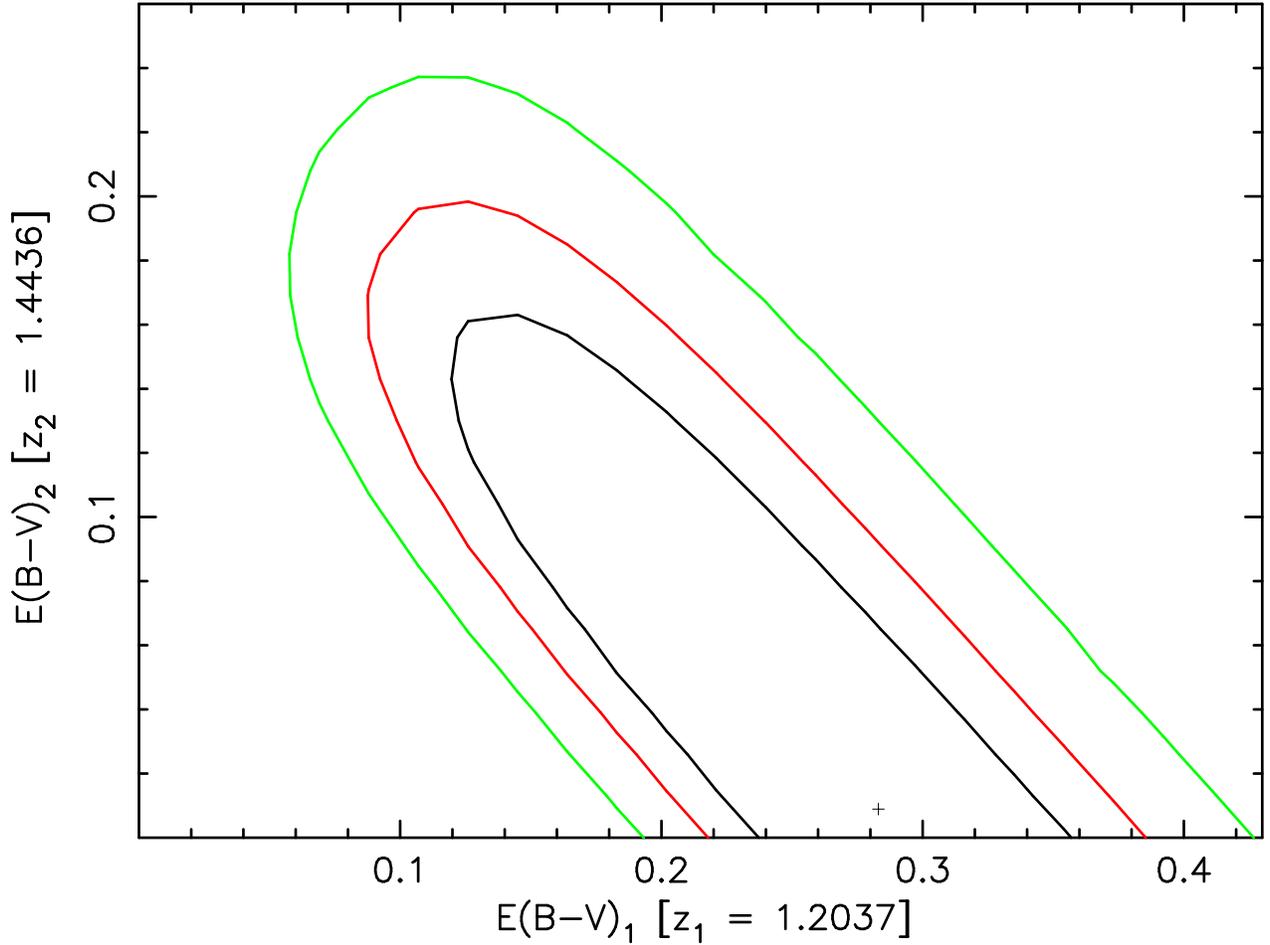}}
\noindent
\end{picture}
\vspace{105mm}
\figcaption{Confidence map in the $E$(B-V)$_1$--$E$(B-V)$_2$ plane. 
The two parameters represent the color correction, assuming $R_V =
2.93$, in two SMC-like dusty complexes at $z_1 = 1.2037$ and $z_2 = 
1.4436$. Contours are 68, 95 and 99.7\% confidence levels and indicate
that a significant fraction of dust must reside in the closer of the
two systems at $z_1$. 
\label{reddening}} 
\end{figure}

An acceptable combined fit of $\chi^2 = 13$ for 25 dof is obtained by
adding SMC-like dust and neutral gas with Magellanic cloud
metallicities ($<$H/Fe$>$ = $-0.5$) at one, free floating
redshift: $\beta = -1.0 \pm 0.1$, $z = 2.8 \pm 0.3$, $E$(B-V) = $0.12
\pm 0.04$ and $\log{\nh} < 2.0 \dex{21}$ cm$^{-2}$.  The alternative
model from Sec.~\ref{sec:spectrum} replaces the $z = 2.8$ gas and dust
with two systems at $z_1$ and $z_2$. Best fit parameters are $\beta =
-1.1 \pm 0.1$, $E$(B-V)$_1 = 0.23 \pm 0.12$, $E$(B-V)$_2 < 0.17$,
$\nh_1 < 1.7 \dex{21}$ cm$^{-2}$ and $\nh_2 < 1.8 \dex{21}$ cm$^{-2}$,
with $\chi^2 = 16$ for 24 dof. These two fits are plotted in
Fig.~\ref{sed}. Plotting the $\chi^2$ landscape of this second model
in the $E$(B-V)$_1$--$E$(B-V)$_2$ plane (Fig.~\ref{reddening}) reveals
that the majority of dust in this scenario is associated with the
closer of the two systems at $z_1$.

In summary, model fits to the spectrum of GRB050318 at T+4,061s where
SMC-like neutral gas and dust are situated at $z_1$ and $z_2$ in front
of a simple powerlaw source reproduce the UVOT spectral index
observed. The spectrum is also well-fit using a single system of gas
and dust at $z = 2.8 \pm 0.3$.

\section{Discussion}
\label{sec:discussion}

\subsection{Dust and gas properties}
\label{sec:dust}

SMC dust has typically proved to be a good fit to extinction curves in
GRB host galaxies, e.g.\ Jakobsson \etal (2004), as one might expect
from a host containing a younger stellar population (Calzetti \etal
2000).  The neutral column density within the host is poorly
constrained, but by adopting the best fit value from the SMC-dust
spectral model with $z = 2.8$, we find a gas-to-dust ratio of
$N$(H{\sc i})/$A$(V) $<$ 7.6 \dex{21} cm$^{-2}$ mag$^{-1}$, $<50$\% of
typical SMC lines-of-sight (Gordon \etal 2003). If we substitute the
SMC dust with the Milky Way prescription of Pei (1992; $R_V = 3.08$)
and assume Solar metallicities in the neutral gas (Anders \& Grevesse
1989) then the fit is also acceptable with $\chi^2$ 12 for 25 dof,
providing slightly different best fit parameters of $\beta = -1.1 \pm
0.1$, $z = 2.4 \pm 0.5$, $E$(B-V) = $0.28 \pm 0.13$ and $\nh < 1.6
\dex{21}$ cm$^{-2}$. The gas-to-dust ratio for this case is $N$(H{\sc
  i})/$A$(V) $<$ 3.0 \dex{21} cm$^{-2}$ mag$^{-1}$, consistent with
the Galactic mean (Bohlin, Savage \& Drake 1978). Similarly, LMC dust
and gas properties with $R_V = 3.16$ (also from Pei 1992) yield $\beta
= -1.0 \pm 0.1$, $z = 2.5 \pm 0.5$, $E$(B-V) = $0.17 \pm 0.07$ and
$\nh < 3.2 \dex{21}$ cm$^{-2}$, $\chi^2$ = 13 for 25 dof. In this
case, $N$(H{\sc i})/$A$(V) $<$ 7.5 \dex{21} cm$^{-2}$ mag$^{-1}$,
which is consistent with the LMC sample from Gordon \etal (2003). This
general model is therefore acceptable statistically for a range of
dust and gas content.

A similar exercise applied to the alternative model, with extinction
and absorption occurring at $z_1$ and $z_2$, yields a poor fit of
$\chi^2$ = 50 for 24 dof when Milky Way dust and gas populations are
assumed. Best fit parameters are $\beta = -1.4 \pm 0.2$, $E$(B-V)$_1 =
0.28 \pm 0.10$, $E$(B-V)$_2 = 0.34 \pm 0.08$, \nh$_1 < 2.0 \dex{21}$
cm$^{-2}$ and \nh$_2 < 2.7 \dex{21}$ cm$^{-2}$.  An LMC gas and dust
model results in $\chi^2$ = 18 for 24 dof, where $\beta = -1.2 \pm
0.1$, $E$(B-V)$_1 = 0.25 \pm 0.15$, $E$(B-V)$_2 = 0.17 \pm 0.12$,
\nh$_1 < 3.2 \dex{21}$ cm$^{-2}$ and \nh$_2 < 2.6 \dex{21}$ cm$^{-2}$.
The gas-to-dust ratio, $N$(H{\sc i})/$A$(V), for the LMC model is
limited to $< 5.0 \dex{21}$ mag$^{-1}$ cm$^{-2}$ in the system at
$z_2$, which is a good candidate for the host galaxy; the same ratio
for the SMC model is unconstrained.  While SMC and LMC models provide
acceptable fits, the Milky Way model does not, although we have made
the simplifying assumptions that the dust contents of the two systems
are identical and that a Ly\,$\alpha$ forest is absent in front of the
burst.  We also note that the constraint from Sec~\ref{sec:sed} that
the majority of dust is located in the $z = 1.2037$ complex can be
dropped if the extinction law in both systems is assumed to be
featureless, e.g. Savaglio \& Fall (2004), where $A(\lambda)/R_V =
E({\mathrm B}-\mathrm{V})(5,500\AA/\lambda)^\delta$. In this scenario
an acceptable fit of $\chi^2 = 16$ for 23 dof is obtained where $\beta
= -1.1 \pm 0.1$, $E$(B-V)$_1 < 0.51$, $E$(B-V)$_2 < 0.46$, $\delta =
1.6^{+1.3}_{-0.8}$, \nh$_1 < 6.4 \dex{20}$ cm$^{-2}$ and \nh$_2 < 4.7
\dex{20}$ cm$^{-2}$.

\subsection{Interpretation}
\label{sec:fireball}

The simplest afterglow emission model assumes synchrotron emission
from a relativistic fireball, expanding into a uniform interstellar
medium (Sari, Piran \& Narayan 1998; Zhang \& M\'{e}sz\'{a}ros
2004). Assuming that the injection break occurs at an energy $< 400$
eV then the X-ray spectral slope of GRB 050318 indicates that $p = 2.4
\pm 0.2$, where $\beta = -p/2$ according to the parameterization of
Sari \etal (1998). The simple fireball model then predicts that the
temporal decay slope should have an index of $\alpha = (2 - 3p)/4 =
-1.3 \pm 0.2$. This is consistent with the XRT decay index and the B
band lower limit, but comparison with the other optical bands is less
convincing. The U and V indices are consistent with $p$ with 14\% and
3\% confidences respectively.

If we assume that the cooling break occurs at an energy greater than
the injection break, and between the UVOT and XRT bandpasses, then the
emission models predicts $\alphauvot = 3(1 - p)/4 = -1.05 \pm 0.2$. In
this case, the U and V indices are consistent with $p$, with
confidences of 80 and 36\% respectively. While the best spectral fit
from Sec.~\ref{sec:sed} does not formerly require a cooling break, it
does not preclude it either. By replacing the model powerlaw continuum
from Sec.~\ref{sec:sed} with a broken powerlaw of fixed spectral
indices $\betauvot = -1.05$ and $\betaxrt = -1.2$, an acceptable fit is
found using SMC dust with $\chi^2 = 13$ for 25 dof, $z =
2.9^{+0.3}_{-0.4}$, $E$(B-V) = $0.15 \pm 0.02$, $\log{\nh} = 19.9 \pm
1.5$ cm$^{-2}$ and providing a lower limit on the cooling break of
$\nu_c > 4.8 \dex{15}$ Hz. The model with two dusty sytems at $z_1$
and $z_2$ yields $\chi^2 = 16$ for 24 dof, $E$(B-V)$_1 = 0.23 \pm
0.12$, $E$(B-V)$_2 < 0.17$, $\log{\nh_1} < 21.1$ cm$^{-2}$,
$\log{\nh_2} = 19.7 \pm 1.5$ cm$^{-2}$ and $\nu_c > 2.4 \dex{15}$ Hz.
While the above model of a slow-cooling fireball within a uniform ISM
fits most of the \swift\ data well in both redshift scenarios, the one
caveat is the inconsistency between the B decay index limit and
$\alpha_{\mbox{\tiny U+V}}$. 

Assuming $z = 2.8$ is the host redshift and a cosmological model of
$H_0 = 65$ \kmsec\ Mpc$^{-1}$, $\Omega_m = 0.3$ and $\Omega_\Lambda =
0.7$, then the BAT 15--350 keV fluence yields an isotropic energy of
$E_{\mathrm{iso}} = (3.6^{+0.7}_{-1.2}) \dex{52}$ erg.  Transforming a
time-averaged value of $E_{\mathrm{p}} = 49$ keV to the rest frame of
the burst, we find $E^\prime_{\mathrm{p}} = 196^{+33}_{-51}$ keV (cf
$E_{\mathrm{iso}} = (1.4^{+0.2}_{-0.4}) \dex{52}$ erg and
$E^\prime_{\mathrm{p}} = 119^{+22}_{-35}$ keV at $z_2 = 1.4436$).
Since both redshifts yield spectral parameters consistent with the
$E_{\mathrm{iso}}$--$E_{\mathrm{p}}$ relationship derived by Amati
\etal (2002), the BAT spectral analysis of this burst cannot provide
useful diagnostics for testing the optical redshift candidates.

\section{Conclusion} 
\label{sec:conclusion}

This paper reports the first significant optical detection of a GRB
afterglow, and subsequent monitoring of the decay, by the UVOT
instrument on-board the \swift\ observatory.  Compared to a simple
powerlaw continuum model, the general deficit of UV emission can be
fit using either gas extinction at redshifts of $z_1 = 1.2037$ and
$z_2 = 1.4426$, corresponding to the absorption systems found by
Berger \etal (2005), or Lyman depletion from an object at $z = 2.8 \pm
0.3$, which would indicate that the two systems at $z_1$ and $z_2$
belong to foreground objects. Consequently the UVOT data cannot
unambiguously determine the host galaxy redshift for this burst. This
will be a common occurence when \swift\ does not detect a UV
source. We note that there is no evidence for a $z = 2.8$ host galaxy
in the $\lambda\lambda$5,000--7,000\AA\ spectroscopy of Berger \etal
(2005), however hosts at this redshift have typically been identified
by absorption features at bluer wavelengths such as a damped
Ly$\alpha$ line (Hjorth \etal 2003) which would occur at 4,617\AA\ at
$z = 2.8$.  Since an identified absorption line provides only a lower
limit to the host redshift, currently available evidence only allows a
lower limit to be placed on the redshift of $z \geq 1.4436$. Decay
curves and the UVOT/XRT SED reveal mostly consistency with the picture
of a slow-cooling fireball in a uniform ISM. However the inferred
steepness of the B band decay slope, relative to U and V, may indicate
some deviations from the simple model.

\acknowledgments  

This work is sponsored at Penn State by NASAs Office of Space Science
through contract NAS5-00136 and at MSSL and Leicester by funding from
PPARC. We gratefully acknowledge the contributions of all members of
the \swift\ team.

\clearpage\newpage

\begin{deluxetable}{rcclc}
\tablecolumns{5}
\tablecaption{UVOT detections ($>2\sigma$ above background) of GRB 050318 
with mid-exposure times relative to the trigger (T+), exposure
durations, filters, AB magnitudes and significance of the detection over 
background. Filter bandpasses are provided in Roming \etal (2005).
\label{tab:mags}}
\tablehead{\colhead{T+ (s)} & \colhead{Exposure (s)} & \colhead{Filter} & 
\colhead{Magnitude} & \colhead{Significance ($\sigma$)}} 
\startdata
3,230 & 100 & V & \hspace{.75em}$17.8^{+0.3}_{-0.2}$ & \hspace{.5em}5.1 \\
3,648 & 100 & U & \hspace{.75em}$19.6^{+0.3}_{-0.2}$ & \hspace{.5em}4.2 \\
5,382 & 880 & B & \hspace{.75em}$18.9^{+0.1}_{-0.1}$ & 19.6 \\
11,201 & 811 & V & \hspace{.75em}$19.1^{+0.2}_{-0.1}$ & \hspace{.5em}6.7 \\
17,041 & 707 & U & \hspace{.75em}$22.0^{+0.6}_{-0.4}$ & \hspace{.5em}2.3 \\
22,827 & 703 & V & \hspace{.75em}$19.5^{+0.3}_{-0.2}$ & \hspace{.5em}4.5 \\
28,609 & 712 & U & \hspace{.75em}$21.8^{+0.5}_{-0.3}$ & \hspace{.5em}2.8 \\
40,193 & 687 & U & \hspace{.75em}$22.1^{+0.8}_{-0.4}$ & \hspace{.5em}2.0 \\
\enddata
\end{deluxetable}

\clearpage\newpage

\clearpage\newpage

\clearpage\newpage

\clearpage\newpage

\end{document}